\documentclass[preprint,superscriptaddress,citeautoscript,floatfix]{revtex4-1}
\usepackage{setspace}
\usepackage{lipsum,booktabs}
\usepackage{amssymb}
\usepackage{amsbsy}
\usepackage{siunitx}
\usepackage{physics}
\usepackage{amsmath}
\usepackage{bm}
\usepackage{textcomp}
\usepackage{gensymb}
\usepackage[breaklinks,colorlinks = true,linkcolor = blue,urlcolor=cyan,citecolor=blue]{hyperref}
\usepackage{multirow}
\usepackage{array}
\usepackage{booktabs}
\usepackage{upgreek}
\usepackage{epsfig,psfrag,amsopn}
\usepackage{mathrsfs}
\usepackage{float}
\usepackage{graphicx}
\usepackage{subcaption}
\usepackage{tabularx}
\usepackage{colortbl}
\usepackage{placeins}
\usepackage[normalem]{ulem}

\begin{document}

\setstretch{1.25}

\title{\small Pressure tuning of structure, magnetic frustration and carrier conduction in Kitaev spin liquid candidate Cu$_2$IrO$_3$: X-ray, Raman, magnetic susceptibility, resistivity and first-principles analysis}

\author{\small Srishti Pal{\color{blue}{$^{\dagger}$}}}
\email[E-mail:~]{srishtipal@iisc.ac.in}
\affiliation{\footnotesize Department of Physics, Indian Institute of Science, Bengaluru 560012, India}

\author{\small Pallavi Malavi{\color{blue}{$^{\dagger , \#}$}}}
\affiliation{\footnotesize Department of Physics, Indian Institute of Science, Bengaluru 560012, India}

\author{\small Arijit Sinha}
\affiliation{\footnotesize Theoretical Sciences Unit, Jawaharlal Nehru Centre for Advanced Scientific Research, Bengaluru 560064, India}

\author{\small  Anzar Ali}
\affiliation{\footnotesize Department of Physical Sciences, Indian Institute of Science Education and Research (IISER) Mohali, Knowledge City, Sector 81, Mohali 140306, India}

\author{\small Piyush Sakrikar}
\affiliation{\footnotesize Department of Physical Sciences, Indian Institute of Science Education and Research (IISER) Mohali, Knowledge City, Sector 81, Mohali 140306, India}

\author{\small Boby Joseph}
\affiliation{\footnotesize Elettra-Sincrotrone Trieste S.C. p.A., S.S. 14, Km 163.5 in Area Science Park, Basovizza 34149, Italy}

\author{\small Umesh V. Waghmare}
\affiliation{\footnotesize Theoretical Sciences Unit, Jawaharlal Nehru Centre for Advanced Scientific Research, Bengaluru 560064, India}

\author{\small Yogesh Singh}
\affiliation{\footnotesize Department of Physical Sciences, Indian Institute of Science Education and Research (IISER) Mohali, Knowledge City, Sector 81, Mohali 140306, India}

\author{\small D. V. S. Muthu}
\affiliation{\footnotesize Department of Physics, Indian Institute of Science, Bengaluru 560012, India}

\author{\small S. Karmakar}
\affiliation{\footnotesize HP$\&$SRPD, Bhabha Atomic Research Centre, Trombay, Mumbai 400085, India}

\author{\small A. K. Sood}
\affiliation{\footnotesize Department of Physics, Indian Institute of Science, Bengaluru 560012, India}

\date{\today}

\begin{abstract}

The layered honeycomb lattice iridate Cu$_2$IrO$_3$ is the closest realization of the Kitaev quantum spin liquid, primarily due to the enhanced interlayer separation and nearly ideal honeycomb lattice. We report pressure-induced structural evolution of Cu$_2$IrO$_3$ by powder x-ray diffraction (PXRD) up to $\sim$17 GPa and Raman scattering measurements up to $\sim$25 GPa. A structural phase transition (monoclinic $C2/c \: \rightarrow$ triclinic $P\bar{1}$) is observed with a broad mixed phase pressure range ($\sim$4 to 15 GPa). The triclinic phase consists of heavily distorted honeycomb lattice with Ir-Ir dimer formation and a collapsed interlayer separation. In the stability range of the low-pressure monoclinic phase, structural evolution maintains the Kitaev configuration up to 4 GPa. This is supported by the observed enhanced magnetic frustration in dc susceptibility without emergence of any magnetic ordering and an enhanced dynamic Raman susceptibility. High-pressure resistance measurements up to 25 GPa in the temperature range 1.4--300 K show resilient non-metallic $R$($T$) behaviour with significantly reduced resistivity in the high-pressure phase. The Mott 3D variable-range-hopping conduction with much reduced characteristic energy scale $T_0$ suggests that the high-pressure phase is at the boundary of localized-itinerant crossover. Using first-principles density functional theoretical (DFT) calculations, we find that at ambient pressure $\rm Cu_2IrO_3$ exists in monoclinic $P2_1/c$ phase which is energetically lower than $C2/c$ phase (both the structures are consistent with experimental XRD pattern). DFT reveals structural transition from $P2_1/c$ to $P\bar{1}$ structure at 7 GPa (involving dimerization of Ir-Ir bonds) in agreement with experimentally observed transition pressure.  

\end{abstract}

\maketitle

\setstretch{1.5}

\section{Introduction}

In the last decade, for possible experimental realizations of Kitaev quantum spin liquid (QSL) ground states~\cite{Kitaev2006,Chaloupka2010}, the quasi-two dimensional (2D) honeycomb iridates and ruthenates have attracted tremendous research interest~\cite{Choi2012,Singh2012,Comin2012,Satapathy2012,Chun2015,Johnson2015,Nasu2016,Nishimoto2016,Williams2016}. In these $d^5$ systems strong spin-orbit coupling (SOC), leading to pseudo spin $J_{eff}$ = 1/2, and edge-sharing octahedral network give rise to anisotropic Ising exchange interaction, the key ingredient of the Kitaev model~\cite{Kitaev2006}. The prime candidates for this model suffer from the concomitant isotropic Heisenberg exchange (often beyond nearest neighbor) and the off-diagonal exchange components (due to non-ideal honeycomb lattice), resulting in undesired long-range magnetic order at low temperature; the zigzag antiferromagnetic (AFM) order in Na$_2$IrO$_3$~\cite{Choi2012} and $\alpha$-RuCl$_3$~\cite{Johnson2015} and incommensurate spiral AFM in $\alpha$-Li$_2$IrO$_3$~\cite{Williams2016}. However, the observation of fractional magnetic excitations based on neutron, Raman and resonant inelastic x-ray scattering measurements have clearly indicated these as proximate quantum spin liquid system~\cite{Nasu2016,Revelli2020,Li2020,Sandilands2015}. Efforts to bring the candidate materials closer to Kitaev limit by suppressing the magnetic order have been successful by isoelectronic cation doping~\cite{Manni2014,Hermann2017} and by application of magnetic field~\cite{Baek2017,Zheng2017,Das2019}. In comparison, similar efforts to tune the various exchange interactions by tuning the crystal structure through application of external pressure have not been successful for these compounds. The 2D honeycomb lattice undergoes distortion even at a moderate pressure leading to Ir-Ir dimer formation~\cite{Clancy2018,Hermann2018,Bastien2018,Hu2018,Li2019}, that either destroys the Ir$^{4+} \: J_{eff}$ = 1/2 configuration~\cite{Clancy2018} or causes magnetic collapse with emergence of a spin-singlet VBS state~\cite{Hermann2018,Bastien2018}. While Na$_2$IrO$_3$, $\alpha$-Li$_2$IrO$_3$ and $\alpha$-RuCl$_3$ exhibit resilient non-metallic character up to very high pressures, the significant drop in resistivity in the high-pressure phases hints for drastic electronic structural modification with possible insulator-metal crossover~\cite{Xi2018,Layek2020,Wang2018}.

In a novel approach to reach Kitaev limit by modifying the honeycomb iridate interlayer bonding has led to the synthesis of a new class of compounds, Cu$_2$IrO$_3$, H$_3$LiIr$_2$O$_6$, and Ag$_3$LiIr$_2$O$_6$~\cite{Abramchuk2017,Kitagawa2018,Bahrami2019}. The interlayer octahedra are replaced by dumbbells in these compounds resulting in $\sim$20-30\% enhanced interlayer separation and more ideal honeycomb configuration. Although a closer to QSL ground state was initially claimed based on the absence of low-temperature magnetic ordering, more careful studies reveal disorder driven formation of random bond singlets in all these compounds resulting in universal scaling behavior in susceptibility and specific heat at low temperatures~\cite{Kitagawa2018,Bahrami2019,Kenney2019,Choi2019,Bahrami2021}. While structural disorder in Ag$_3$LiIr$_2$O$_6$ is synthesis related with a scope to improve~\cite{Bahrami2021}, interlayer proton disorder in H$_3$LiIr$_2$O$_6$~\cite{Knolle2019,Yadav2018,Li2018} and presence of mixed valent Cu$^+$/Cu$^{2+}$ within intra-layer CuO$_6$ octahedra~\cite{Kenney2019,Choi2019} are of fundamental nature. In spite of bond-disordering, the observed fractional excitations in Raman measurements and NQR study have indicated proximate Kitaev QSL behavior in the latter two compounds~\cite{Pal2021,Takahashi2019,Pei2020}. More recently, optical phonons renormalization below the Kitaev temperature scale of $\sim$120 K has been quantitatively attributed to the interaction of phonons with the Majorana fermions~\cite{Pal2021}. Recent high-pressure studies on the Cu$_2$IrO$_3$ system~\cite{Fabbris2021,Jin2022} have detected two pressure-induced structural transitions at $\sim$7 GPa and 11-15 GPa. While the first transition is reported to be to a triclinic $\alpha-P\bar{1}$ followed by Ir-Ir dimerization, the high-pressure structure after the second transition was only predicted by DFT based evolutionary structure search method and could not be uniquely determined by a successful refinement of the PXRD pattern. Interestingly, high-pressure resistivity measurements on this system~\cite{Jin2022} have shown that the second structural transition is associated with an insulator to metal transition. However, DFT calculations~\cite{Fabbris2021} could not capture the same and predicted opening up of a gap of $\sim$0.3 eV in the dimerized phase. Hence, the effect of hydrostatic pressure in this proximate Kitaev QSL candidate is yet to be examined thoroughly, thus motivating this present study.

We report here the structural, vibrational, magnetic, and electronic properties of Cu$_2$IrO$_3$ up to $\sim$25 GPa using a combination of PXRD, Raman scattering, magnetic susceptibility, resistivity measurements, and density functional theoretical calculations. A structural phase transition (monoclinic $C2/c \: \rightarrow$ triclinic $P\bar{1}$) is observed in a broad mixed-phase pressure range ($\sim$4 to 15 GPa). The triclinic phase consists of heavily distorted honeycomb lattice with Ir-Ir dimer formation and a collapsed interlayer separation. In the stability range of the low-pressure monoclinic phase (up to $\sim$4 GPa), structural evolution with pressure drives the system marginally towards the Kitaev limit, supported by the observed enhanced magnetic frustration in dc susceptibility without emergence of any magnetic ordering and an enhanced dynamic Raman susceptibility. High-pressure resistance measurements show resilient non-metallic $R$($T$) up to 25 GPa, but with significantly reduced resistivity in the high-pressure phase. The Mott 3D variable-range-hopping conduction with much reduced characteristic energy $T_0$ indicates enhanced density of state at Fermi level suggesting the high-pressure phase to be at the boundary of localized-itinerant crossover. Our DFT calculations capture a pressure dependent structural transition at $\sim$7 GPa, but from $P2_1/c$ to $P\bar{1}$ structure, which involves dimerization of Ir-Ir bonds in 2D honeycomb layers of Cu$_2$IrO$_3$. We show that presence of mixed valence disorder in Ir-Cu honeycomb layers are responsible for theoretically predicted~\cite{Fabbris2021} metallicity in Cu$_2$IrO$_3$ at ambient pressure. At ambient pressure, Cu$_2$IrO$_3$ is an insulator having an energy band gap of 0.3 eV (as extracted from the high $T$ Arrhenius fit of our resistivity data), which has been simulated by substituting Cu in honeycomb layers with Li atoms. While our theoretically obtained optimized structure of Cu$_2$IrO$_3$ at higher pressures ($P\bar{1}$ phase) differs from experimentally observed high-pressure structure having higher degree of distortions and enhanced dimerization, our calculated frequencies of Raman-active phonon modes at different pressures qualitatively agrees with observed pressure dependent Raman spectra.

\section{Experimental and Computational Details}

High quality polycrystalline samples of Cu$_2$IrO$_3$ were prepared by topotactic cation exchange reaction of Na$_2$IrO$_3$ with CuCl. The samples are from the same batch as used in recent temperature-dependent Raman study~\cite{Pal2021} and the results of detailed characterization by magnetic susceptibility and muon spin resonance are reported previously~\cite{Choi2019}.

High-pressure powder X-ray diffraction measurements on Cu$_2$IrO$_3$ at room temperature have been performed at the XPRESS beam line ($\lambda$ = 0.49585 \AA) of the Elettra Synchrotron, Trieste, Italy. Two measurement runs were performed by mounting finely ground sample in a membrane-driven symmetric diamond anvil cell (DAC)-- (i) run1 up to 10.7 GPa in finer pressure steps with methanol-ethanol-water (MEW) (16:3:1) as pressure transmitting medium (PTM) and (ii) run2 up to 17 GPa in larger steps with silicone oil as PTM. The 2D diffraction images were recorded on a DECTRIS PILATUS3 S 6M detector and these were converted to \textit{I} vs 2$\theta$ diffraction profiles using the Dioptas software~\cite{Prescher2015}. Rietveld refinements were performed for obtaining the high-pressure structural parameters using EXPGUI software~\cite{Toby2015}. For high-pressure Raman experiments, a small piece ($\sim$150 micron) of Cu$_2$IrO$_3$ was loaded inside a Mao Bell type DAC. Solid NaCl crystal was used as the PTM rather than the ethanol+methanol mixture as the latter gave an enhanced low frequency background. Raman spectra were recorded at room temperature in a backscattering geometry using Horriba LabRAM HR Evolution Spectrometer with a DPSS laser of wavelength 532 nm and using 1.5 mW laser power. The spectra were collected by a thermoelectric cooled charge coupled device (CCD) (HORIBA Jobin Yvon, SYNCERITY 1024 $\times$ 256).

High-pressure resistance measurements at low temperatures have been performed on a pressed pellet of the polycrystalline sample of $\sim$10 $\mu$m thick and $\sim$100 $\mu$m lateral dimensions. The resistance was measured using standard (quasi-) four probe method with suitable excitation current (depending on resistance range) and with ac lock-in detection technique. A Stuttgart version DAC was used for measurements under quasi-hydrostatic pressures up to 25 GPa using finely powdered NaCl as the PTM. The DAC was placed inside a KONTI-IT (Cryovac) cryostat. The high-pressure magnetic susceptibility ($\chi$) in a \textit{dc} field of 10 mT was measured in a SQUID magnetometer (S700X Cryogenic Ltd.) using nonmagnetic Cu-Be cells; a piston-cylinder cell (CamCool) for measurements up to 1.1 GPa and a DAC (Mcell-ultra from EasyLab) for measurements up to $\sim$4 GPa, with daphne oil as PTM. The thermo-magnetic responses of the empty pressure cells were measured separately and subtracted from corresponding data. The pressures inside the DACs were measured by conventional ruby luminescence method. For $\chi$ measurements in piston-cylinder cell, pressure was determined from the known $P$-dependent superconducting $T_c$ variation of Sn~\cite{Eiling1981}.

Our first-principles calculations are based on density functional theory (DFT) as implemented in VASP package~\cite{PhysRevB.47.558, KRESSE199615, PhysRevB.54.11169} and interaction between valence electrons and ions was modelled using projector-augmented-wave (PAW) pseudopotentials~\cite{PhysRevB.50.17953, PhysRevB.59.1758}. Perdew-Bruke-Ernzerhof (PBE) parametrization of exchange correlation energy functional within a Generalized-gradient approximation (GGA) was used for treating exchange-correlation energy of electrons~\cite{PhysRevLett.77.3865, PhysRevLett.78.1396}. Energy cut-off of 500 $eV$ was used to truncate the plane wave basis-sets used to represent Kohn-Sham (KS) wave functions. Brillouin zone (BZ) integrations were sampled on a uniform mesh of $4 \times 2 \times 4$ Monkhorst-pack $k$-points~\cite{PhysRevB.13.5188} with a Gaussian smearing width of $k_BT$ = 0.04 eV. Self-consistent numerical solution of the Kohn-Sham equation was obtained iteratively to converge total energy within 10$^{-6}$ eV. Structures were relaxed to minimize total energy, until Feynman-Hellman forces on each atom are less than 10$^{-3}$ eV/\AA.

In these calculations we included effects of spin orbit coupling (SOC) and onsite correlation with Hubbard-$U$~\cite{Anisimov_1997} parameters ($U_{eff}$) 2 eV and 6.5 eV for Ir $5d$ and Cu $3d$ orbitals respectively~\cite{Fabbris2021}. Dynamical matrix and phonon spectrum at $\Gamma$-point were obtained using frozen phonon method implemented in PHONOPY package~\cite{phonopy}. SPGLIB package~\cite{togo2018textttspglib} was used extensively to determine symmetry groups of the structures, SOC was not included in phonon calculations to reduce computation time.

All the structures were visualized using VESTA~\cite{Momma:db5098} package, Brillouin zone were drawn using XCrySDen package~\cite{KOKALJ1999176} and ISODISPLACE package~\cite{Campbell:wf5017} was used to find atomic displacements in the structure during structural phase transformation.

\section{Results and Discussion}
\subsection{X-ray Diffraction}

\begin{figure}[h!]
\centering
\includegraphics[width=160mm,clip]{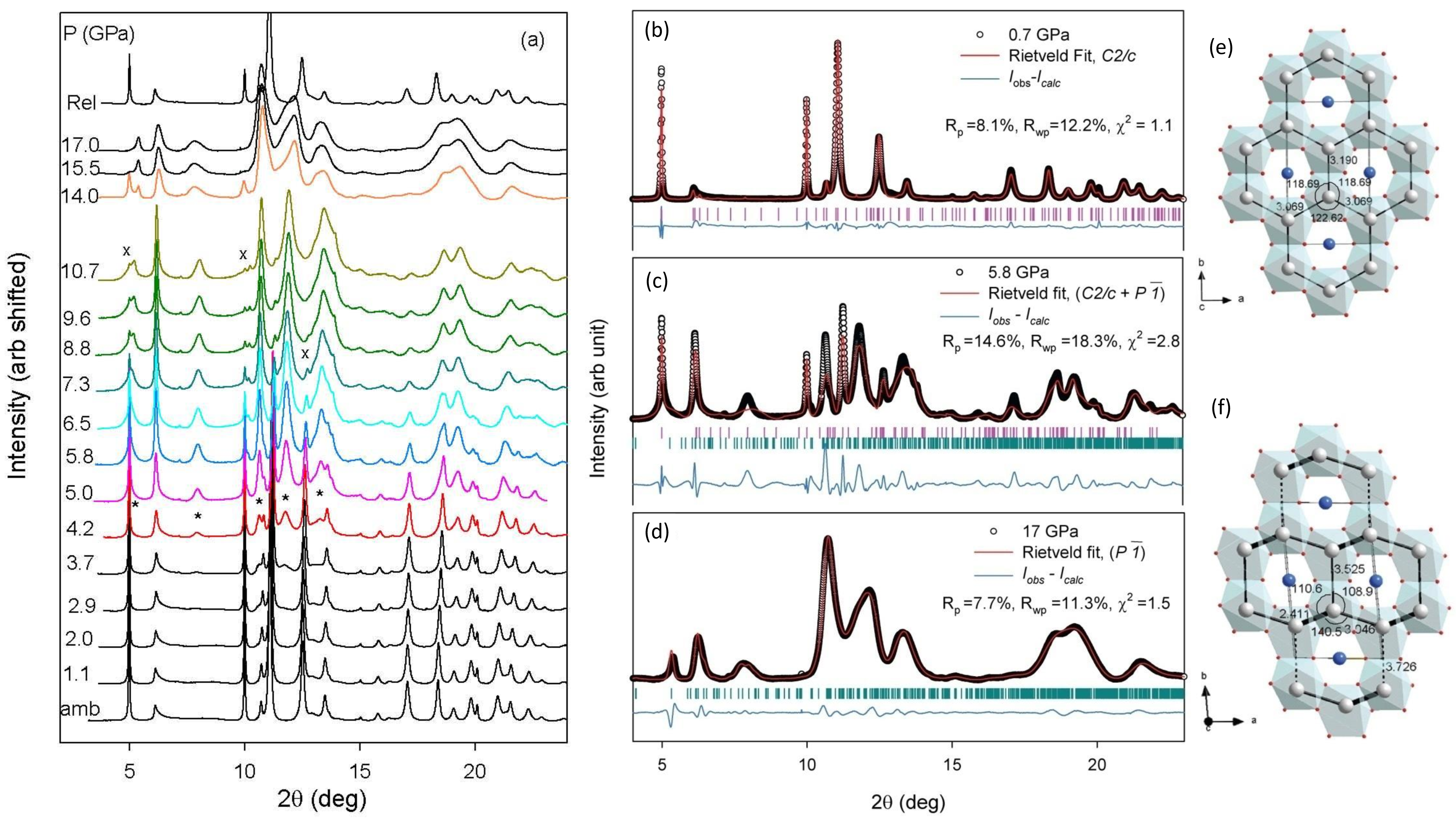}
\caption{\small (a) X-ray diffraction patterns (intensity arbitrarily shifted) of Cu$_2$IrO$_3$ at various high pressures from two different runs (plots up to 10.7 GPa are with MEW as PTM and displayed patterns of higher pressures are with silicone oil as PTM). Rietveld refinement of the X-ray diffraction profiles of Cu$_2$IrO$_3$ (b) at 0.7 GPa with $C2/c$ structure (c) at 5.8 GPa with mixed phases, $C2/c$ and $P\bar{1}$ structure and (d) at 17 GPa with $P\bar{1}$ structure. (e) Monoclinic ($C2/c$) structure honeycomb layer at 0.7 GPa (f) Distorted honeycomb layer in triclinic $P\bar{1}$ structure at 17 GPa. The Ir-Ir bond lengths and Ir-Ir-Ir angles are labelled for both structures.}
\label{Fig1}
\end{figure}

High-pressure X-ray diffraction measurements at room temperature show a structural phase transition of the quasi-2D honeycomb lattice compound Cu$_2$IrO$_3$. Ambient monoclinic structure (SG: $C2/c$, $z$ = 8) is found to be stable up to $\sim$4 GPa, beyond which emergence of several new Bragg peaks [shown by `*' in Fig.~\ref{Fig1}(a)] indicate occurrence of a structural phase transition, with low pressure monoclinic structure persisting up to $\sim$15 GPa. Also, much reduced Bragg peaks intensities of the low-pressure monoclinic phase are shown by `$\times$' at 7.3 and 10.7 GPa in Fig.~\ref{Fig1}(a). Mixed phase patterns are shown in colour. The structural transition is found completely reversible, as shown in top profile of Fig.~\ref{Fig1}(a) at $\sim$0.1 GPa released from 17 GPa. Based on the diffraction pattern at 17 GPa where monoclinic phase peaks disappear completely, the high-pressure structure is indexed as triclinic (SG: $P\bar{1}$, $z$ = 8) where atomic positions are estimated using the PowderCell software~\cite{Kraus1996} and further refined by Rietveld method~\cite{Toby2015}. Figure~\ref{Fig1} shows the Rietveld fit of the diffraction patterns in three pressure ranges: (b) monoclinic $C2/c$ phase below 4 GPa, (c) mixed phase of monoclinic $C2/c$ and triclinic $P\bar{1}$ in the range of 4 to 15 GPa and (d) triclinic ($P\bar{1}$) phase above 15 GPa. Because of the broadened and overlapped peaks in the triclinic phase, the mixed phase Rietveld analysis does not result in good fit (as apparent from the obtained higher $R_{wp}$). Therefore, for more accurate lattice parameter determination, Le-Bail profile fitting analyses were performed without attempting to refine atomic positions. While in the low $P$ monoclinic structure the 2D honeycomb lattice remains nearly isotropic [Fig.~\ref{Fig1}(e)], this becomes quite distorted in the high-pressure triclinic structure [with four different Ir-Ir bond lengths in the unit cell, the shortest one being $\sim$2.41 \AA~as shown in Fig.~\ref{Fig1}(f)] with possible formation of Ir-Ir covalent dimers. An enhanced structural disorder in the high-pressure triclinic phase is also apparent from the intrinsic broadened Bragg peaks compared to the monoclinic phase. The low angle ($2\theta = 5^{\circ}$) single Bragg peaks (002) of both the phases help identifying the mixed phase range and also the c-axis collapsed nature of the triclinic structure~\cite{suppl}. Moreover, a strongly enhanced intensity of the 2$\theta$ = 6$^{\circ}$ peak is a clear signature of heavily distorted honeycomb lattice in the triclinic phase. However, due to the inability to refine the O-positions in the structural analysis of the high-pressure phase, we cannot comment on the distortion (bend/buckling) of the intra-layer IrO$_6$ octahedra and interlayer O-Cu-O dumb-bells in the high \textit{P} structure.

\begin{figure}
\centering
\includegraphics[width=120mm,clip]{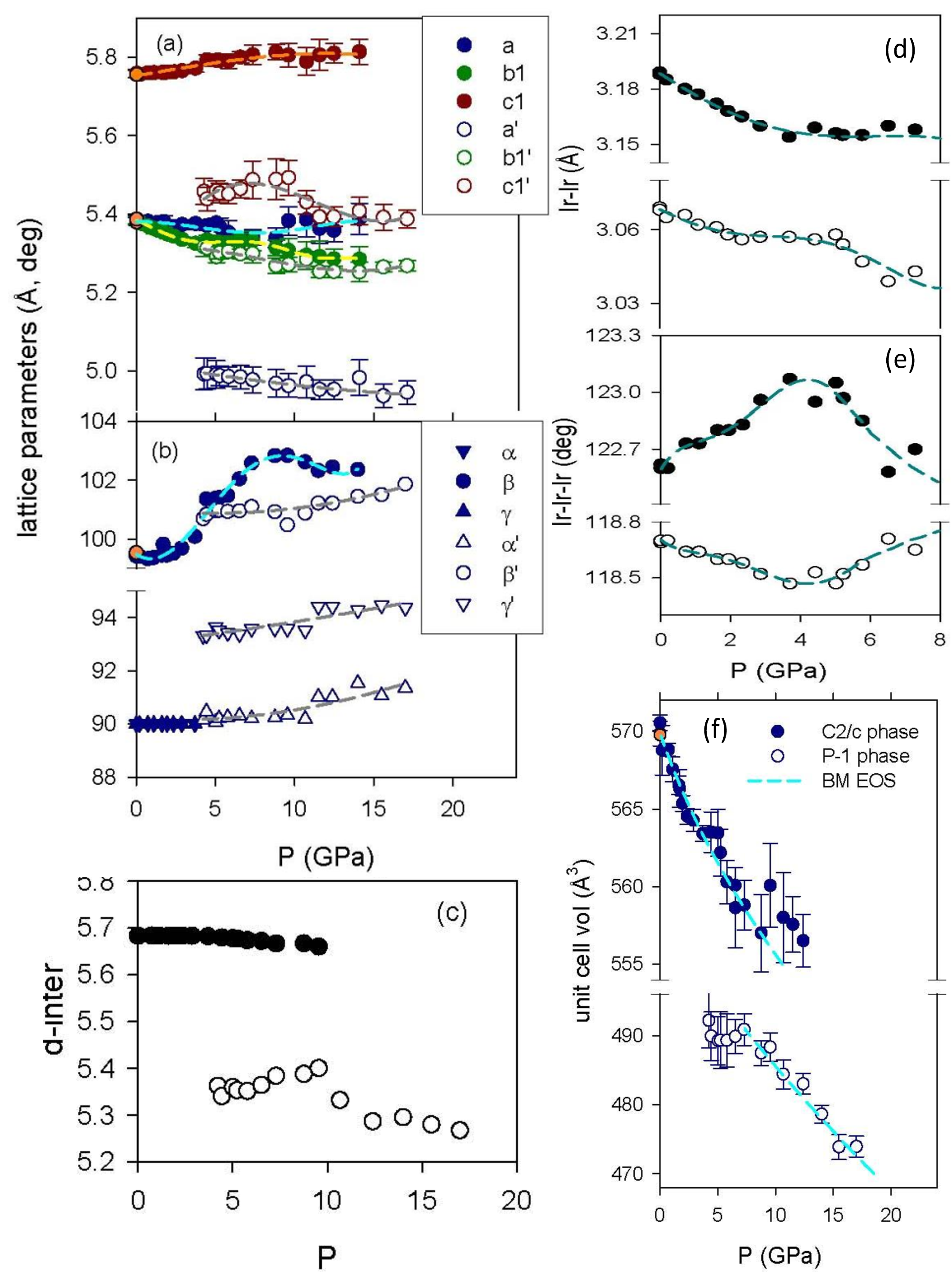}
\caption{\small (a)-(b) Pressure dependence of the lattice constants across the structural transition. Solid symbols are for low pressure monoclinic structure ($C2/c$) and open symbols (also indicated by prime legends) for high-pressure triclinic structure ($P\bar{1}$). Dashed lines are guides to eye for each plot. For both structures, $b1 = b/\sqrt{3}$ and $c1 = c/2$. (c) Interlayer (honeycomb) separation $d$-inter is plotted as a function of $P$. Variations of (d) Ir-Ir bond distances and (e) Ir-Ir-Ir bond angles in the low \textit{P} monoclinic structure with pressure. (f) Unit cell volume as a function of pressure in across the structural transition. Dashed cyan lines are the Birch-Murnaghan equation of state fit for both the phases (see text for details).}
\label{Fig2}
\end{figure}

In Fig.~\ref{Fig2}(a) and (b) are plotted the changes in lattice parameters of both phases as a function of pressure. Although the monoclinic $c1$ parameter ($c1 = c_{mono}/2$) increases monotonically upon increasing pressure, the interlayer separation remains mostly unchanged up to 4 GPa due to the increased $\beta$ angle and decreases gradually at higher pressures~\cite{suppl}. The collapsed interlayer spacing in the high-pressure triclinic phase is apparent. The in-plane lattice parameters ($a = a_{mono}$, $b1 = b_{mono}/\sqrt{3}$) show anisotropic compression resulting in increased distortion in the honeycomb lattice up to 4 GPa which reduces anomalously at higher pressures leading to a large distortion above 8 GPa. The monoclinic $\beta$ angle increases monotonically up to 8 GPa and marginally decreases at higher pressures. As shown in Fig.~\ref{Fig2}(c) and (d), the Ir-Ir bond distances decrease by $\sim$1\% up to 8 GPa, whereas the Ir-Ir-Ir bond angles increase by 0.2 - 0.4$^{\circ}$ up to 4 GPa and then reach to ambient pressure values at $\sim$9 GPa. This implies that the high-pressure monoclinic phase is close to the regular Kitaev honeycomb lattice with reduced Ir-Ir bond distances. This along with unchanged inter-layer separation results in octahedral elongation along the trigonal axis ($\perp$ to the $ab$ plane) bringing this towards $O_h$ symmetry.

High-pressure triclinic phase ($P\bar{1}$) can be identified at 4.2 GPa, although with a low phase fraction~\cite{suppl} causing its lattice parameters determination with large uncertainty [Fig.~\ref{Fig2}(a)-(b)]. At pressures above 9 GPa, with increasing phase fraction, the lattice parameters show systematic pressure dependence. The structural transition is found to be completely reversible and the observed volume discontinuity between two phases at 9 GPa is $\sim$12\%. The structural details of the ambient monoclinic phase and high-pressure triclinic phase at 17 GPa and at 5.8 GPa are listed in Table \textcolor{blue}{S1}~\cite{suppl}.

\subsection{Raman Scattering Results}

\begin{figure}
\centering
\includegraphics[width=140mm,clip]{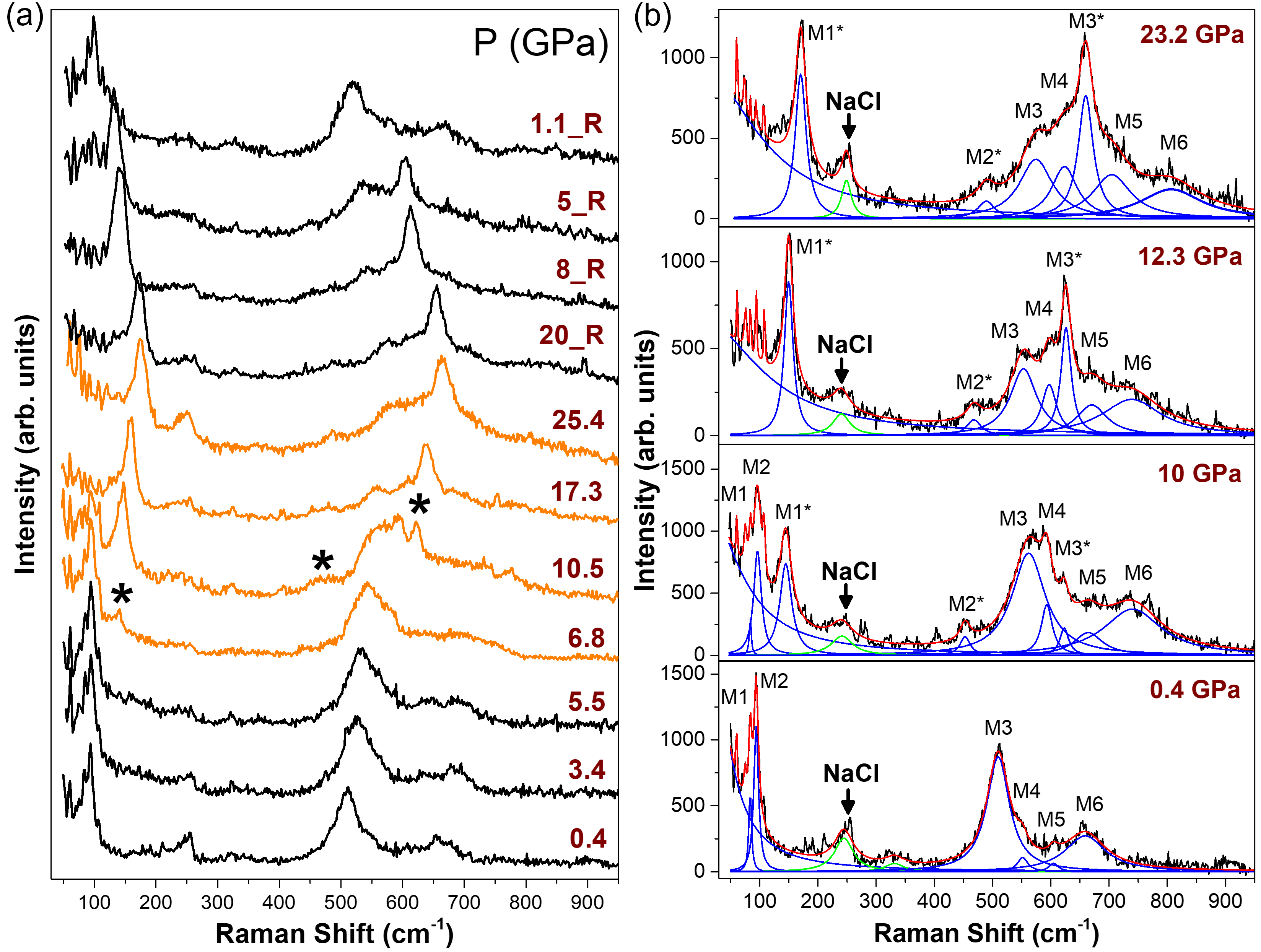}
\caption{\small (a) Raman spectra of Cu$_2$IrO$_3$ at selected pressure values for pressurizing to 25.4 GPa and depressurizing (denoted by R) back to 1.1 GPa. Asterisks (*) denote the new peaks emerging at the monoclinic to triclinic transition. (b) Phonon fits to the Raman profile at selected pressures. Black curves denote the experimental data. Individual phonon modes and the cumulative fits are shown by blue and red curves, respectively. The green curves are the NaCl peaks (used as the PTM). M1*-M3* denote the new modes appearing in the high $P$ structure over the existing ones (M1-M6).}
\label{Fig3}
\end{figure}

The vibrational spectra of Cu$_2$IrO$_3$ for ambient $C2/c$ monoclinic symmetry, is comprised of 39 Raman active $\Gamma$-point phonon modes, $\Gamma_{Raman}$ = 18$A_g$ + 21$B_g$. Among these six Raman modes at 84, 94, 510, 551, 605, and 659 cm$^{-1}$ [marked M1-M6 in Fig.~\ref{Fig3}(b)] along with a quasi-elastic scattering (QES) background below 150 cm$^{-1}$ were observed in our ambient Raman spectrum.

Figure~\ref{Fig3}(a) portrays the pressure evolution of the Raman spectra of Cu$_2$IrO$_3$ while pressurizing up to 25.4 GPa and upon decompression to 1.1 GPa. All the phonon modes including the QES signal have been fitted with Lorentzian line-shapes for the entire pressure range as shown in Fig.~\ref{Fig3}(b). The similar local symmetry of the edge-sharing RuCl$_6$ or IrO$_6$ octahedra of the honeycomb planes in neighbour Kitaev compounds $\alpha$-RuCl$_3$~\cite{Li2019} and $\alpha$-Li$_2$IrO$_3$~\cite{Li2020} allows qualitative assignment of the Raman modes in Cu$_2$IrO$_3$. While the low-frequency M1, M2 modes are primarily related to the vibration of the in-plane Ir network, the high-frequency M3-M6 modes correspond to the vibration of the IrO$_6$ octahedra including Ir-O-Ir-O ring breathing, Ir-O-Ir-O plane shearing, and breathing of the upper and lower oxygen layers.

The ambient Raman spectral features remain unchanged under pressure up to $\sim$6 GPa with expected blueshift of the phonon modes beyond which the spectra show pronounced changes. At 6.8 GPa, a new mode at 140 cm$^{-1}$ (marked as M1*) emerges near the low frequency mode M2, while a simultaneous significant change is also noticed in the high frequency band. The weak shoulder peak M4 at 551 cm$^{-1}$ starts gaining intensity significantly above this pressure. This is followed by the appearance of other two modes, one weak mode at 457 cm$^{-1}$ (M2*) and other at 622 cm$^{-1}$ (M3*) (clearly visible above 10 GPa), and disappearance of the M1 and M2 modes above $\sim$11 GPa. Appearance of new modes clearly indicates a structural transition to a lower symmetry, supporting the monoclinic to triclinic transition evidenced by our high-pressure XRD results.

\begin{figure}
\centering
\includegraphics[width=80mm,clip]{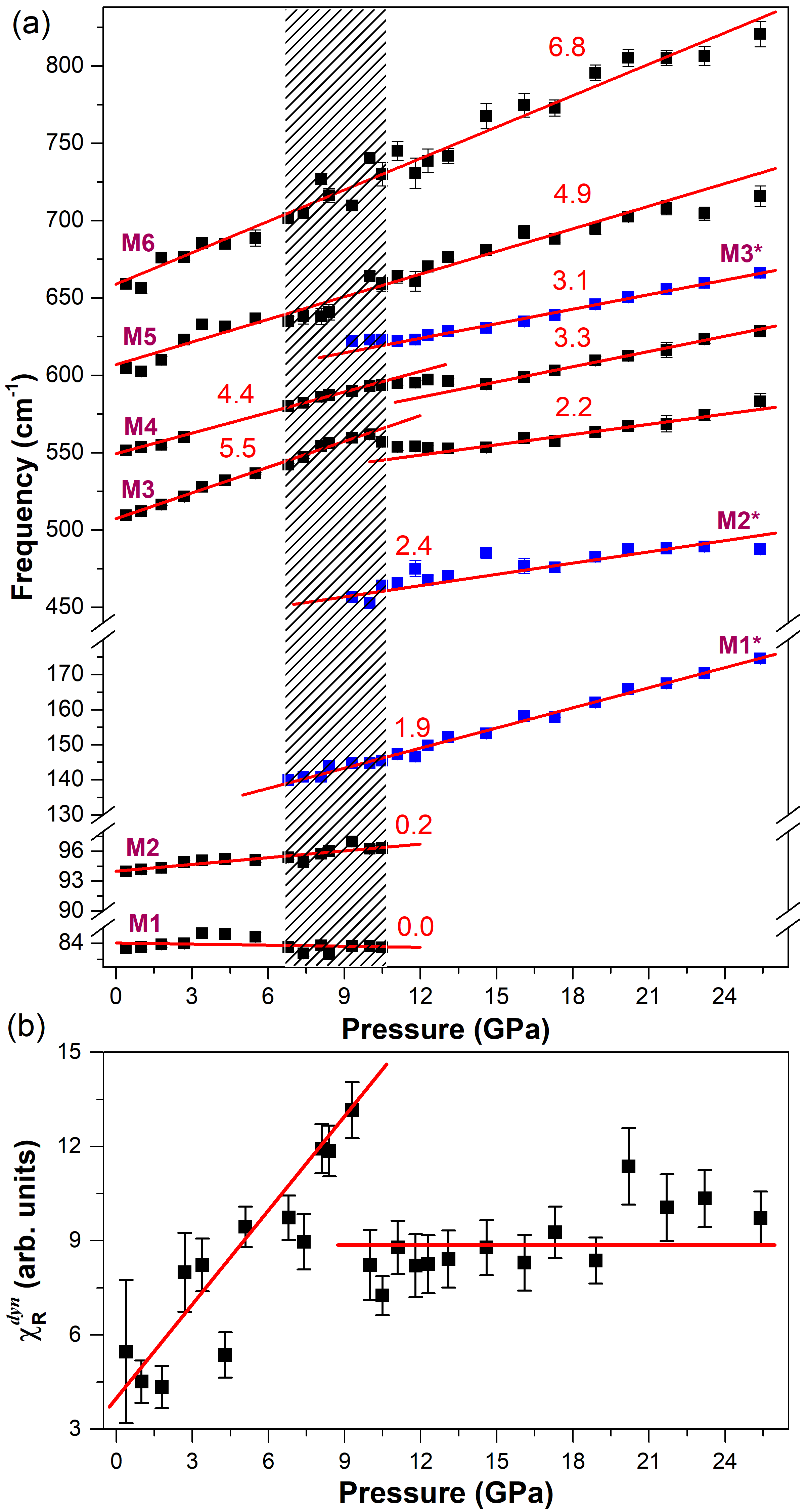}
\caption{\small (a) Pressure evolution of phonon frequencies of Cu$_2$IrO$_3$. Black squares denote the phonon modes of the low $P$ monoclinic phase. Blue squares are the new modes appearing at the onset of the structural transition. Red solid lines are the linear fits of phonon frequencies with pressure ($\frac{d\omega}{dP}$ values mentioned). (b) Pressure dependence of the dynamic Raman susceptibility $\chi_R^{dyn}$. Red solid lines in are guide to eyes.}
\label{Fig4}
\end{figure}

The emergence of the M1* mode (140 cm$^{-1}$) at a higher frequency than that of M1 and M2, indicates considerably reduced intra-layer Ir-Ir bond length, consistent with the high-pressure triclinic crystal structure. The emergence of the M3* mode and M4 mode gaining intensity with pressure can be due to lifting of the degeneracy of the IrO6 vibrations due to octahedral distortion and/or symmetry breaking by the dimer formation. Upon decompression from 25 GPa, the Raman spectra regains its ambient features [Fig.~\ref{Fig3}(a)], showing reversibility of the structural transition. Figure~\ref{Fig4}(a) depicts the pressure evolution of the mode frequencies along with linear fits. All modes exhibit expected blueshift in frequency with increasing pressure due to reduction in the unit cell volume. The M3 and M4 modes clearly show that the slopes, $\frac{d\omega}{dP}$, change across the structural transition. The phonon frequencies ($\omega$), $\frac{d\omega}{dP}$ values, and the corresponding Gr\"{u}neisen parameters $\gamma_i$ = $\frac{B_0}{\omega}\:\frac{d\omega}{dP}$ of both the phases are listed in Table \textcolor{blue}{S2}~\cite{suppl}. The bulk moduli, $B_0$ = 203(8) GPa for low pressure monoclinic phase, and $B_0$ = 208(13) GPa for the high $P$ phase, are obtained from our high-pressure XRD results [see Fig.~\textcolor{blue}{S3}]~\cite{suppl}. The shaded region of $\sim$7-11 GPa shown in Fig.~\ref{Fig4}(a) represents the presence of the mixed phase where the new modes (M1*-M3*) develop gradually and above which the old M1 and M2 modes disappear. A slightly different mixed phase pressure range compared to that ($\sim$4-15 GPa) in XRD results can be attributed to different sensitivities of two probes.

Quantum spin systems with strong spin-orbit coupling often give rise to significant light scattering from the fluctuations in the spin density~\cite{Reiter1976,Lemmens2013} resulting in quasi-elastic Raman response. The dynamic response from the spin system can be quantified in terms of the dynamic Raman susceptibility $\chi_R^{dyn}$ obtained by integrating the Raman conductivity $\frac{\chi''(\omega)}{\omega}$ using the Kramers-Kronig relation
\begin{equation}
\chi_R^{dyn} = \lim_{\omega \to 0} \chi(\omega, k = 0) = \frac{2}{\pi} \int \frac{\chi''(\omega)}{\omega} d\omega
\label{eq_KK}
\end{equation}
The dynamic Raman tensor susceptibility $\chi''(\omega) = \frac{I(\omega}{n(\omega)+1}$ ($n(\omega)$ is the Bose thermal factor) represents the imaginary part of the general susceptibility tensor $\chi(r,t)$~\cite{Cottam1986}.

Figure~\ref{Fig4}(b) illustrates the pressure dependence of the dynamic spin susceptibility $\chi_R^{dyn}$ of Cu$_2$IrO$_3$ with integration carried out from 50-1000 cm$^{-1}$ after subtracting the phonon spectra. $\chi_R^{dyn}$ increases with pressure in the monoclinic phase (up to $\sim$9 GPa). We speculate that pressure-induced structural evolution towards a regular Kitaev configuration in the monoclinic phase (as discussed below) increases the fluctuations in the exchange coupled Ir spin sites resulting in the enhancement of $\chi_R^{dyn}$. The significant drop in $\chi_R^{dyn}$ at higher pressures can be attributed to the reduced fluctuations (released frustration) in the distorted high-pressure triclinic structure.

\subsection{Magnetic Susceptibility}

\begin{figure}
\centering
\includegraphics[width=80mm,clip]{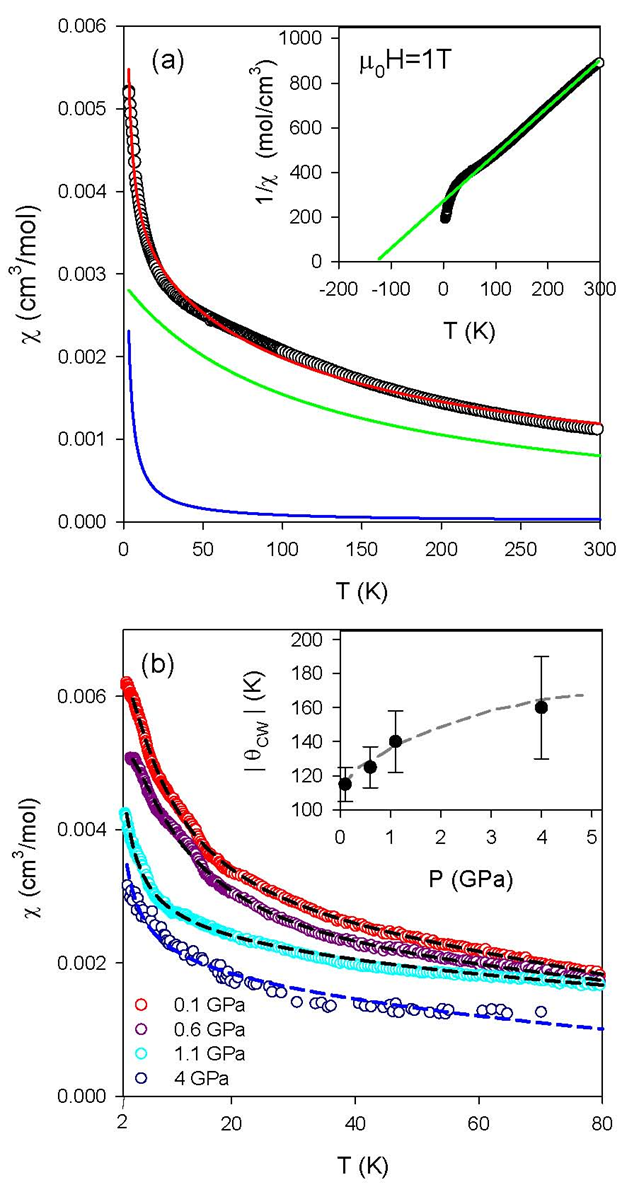}
\caption{\small (a) Zero-field-cooled \textit{dc} susceptibility measured at 1 T field. The $\chi$($T$) curve has been fitted (red solid line) with combination of Curie-Weiss term (green line) and a Curie term (blue line) (see text). Inset, 1/$\chi$ vs $T$ plot with linear extrapolation of 1/$\chi$ from high $T$ (shown by green line) showing deviation from the CW law at low temperatures. \textbf{(b)} Field cooled susceptibility (at 10 mT field) plots as a function of temperature up to 4 GPa. The $\chi$($T$) curves at high pressures have been fitted by two terms as in (a). The pressure variation of the CW temperature in the first term is shown as inset.}
\label{Fig5}
\end{figure}

Figure~\ref{Fig5}(a) shows zero-field-cooled (ZFC) \textit{dc} susceptibility ($\chi$) at ambient pressure measured using 1 T field. The $\chi$(\textit{T}) fits well over the entire temperature range with a Curie-Weiss (CW) term and a Curie term (that becomes prominent at low temperature),
\begin{equation}
\chi(T) = \chi_0 + \frac{C}{(T - \Theta_{CW})} + \frac{C_{imp}}{T}
\label{eq_CW}
\end{equation}

The CW behaviour indicates dominant AFM interaction ($\Theta_{CW}$ = -115 K and $\mu_{eff}$ = 1.63 $\mu_B$), while the Curie term due to defect spins becomes prominent below $\sim$20 K. Considering that the defect spins in Cu$_2$IrO$_3$ are due to presence of mixed valent Cu$^+$/Cu$^{2+}$ in the honeycomb layer (and consequently affecting a neighbouring Ir atom valency and its spin configuration), it is believed that these spins may not be free (unlike the interlayer non-interacting spins in Herbertsmithite resulting from antisite disorder~\cite{Malavi2020}). Low temperature $\chi$($T$) was thus shown previously to follow a sub-Curie fit ($\chi = \frac{C_{imp}}{T^{\alpha}}$, $\alpha$ = 0.26), that was explained by bond-disordered QSL state (by the formation of random spin-singlets) as a result of the above quenched disorder~\cite{Choi2019}. Presence of defect spins at low $T$ in our sample is also evident from the observed non-linear $M$($H$) at low temperatures~\cite{suppl}.

Magnetic properties up to 4 GPa (in the stability range of the $C2/c$ phase) have been investigated in two arrangements: a piston-cylinder pressure cell (with 20 mg sample) for studies up to 1.1 GPa and a miniature diamond anvil cell for studies at higher $P$. At each pressure, empty cell background moment was measured and subtracted carefully to obtain the sample magnetic moment. No long-range magnetic order has been detected below 100 K at high-pressures. In Fig.~\ref{Fig5}(b), we plot the \textit{DC} susceptibility in the $T$ range 2-100 K at various pressures up to 4 GPa. At higher $P$, the overall $\chi$ gets suppressed systematically. The $\chi$($T$) data, fitted by Eq.~\ref{eq_CW}, shows a pressure-induced systematic increase of the CW temperature [inset of Fig.~\ref{Fig5}(b)]. In absence of any magnetic ordering, this indicates that spin frustration increases with pressure in the low-pressure monoclinic phase resulting from the structural modification, as discussed later (in agreement with systematic increase of Raman dynamic susceptibility).

\subsection{Electrical Transport Measurements}

To understand the effect of pressure on the electronic properties of Cu$_2$IrO$_3$ we have performed resistance measurements up to 25 GPa. In order to remove possible resistance contribution of grain boundaries, we measured room temperature resistance in three pressure cycles up to 12 GPa. While the $R$($P$) variation in first and second cycle differ significantly, this remained almost unchanged in second and third cycle, indicating negligible grain boundary contribution in the pressure cycled sample (as shown in Fig. \textcolor{blue}{S6})~\cite{suppl}. Since at low pressures the resistance increases by six orders of magnitude at low $T$, a quasi four-probe measurement was performed below $\sim$8 GPa and standard four probe at higher pressures to improve the signal-to-noise at all pressures. The $R$($T$) data at each pressure were collected in slow warming cycle ($\sim$1 K/min) to avoid thermal drift.

\begin{figure}
\centering
\includegraphics[width=140mm,clip]{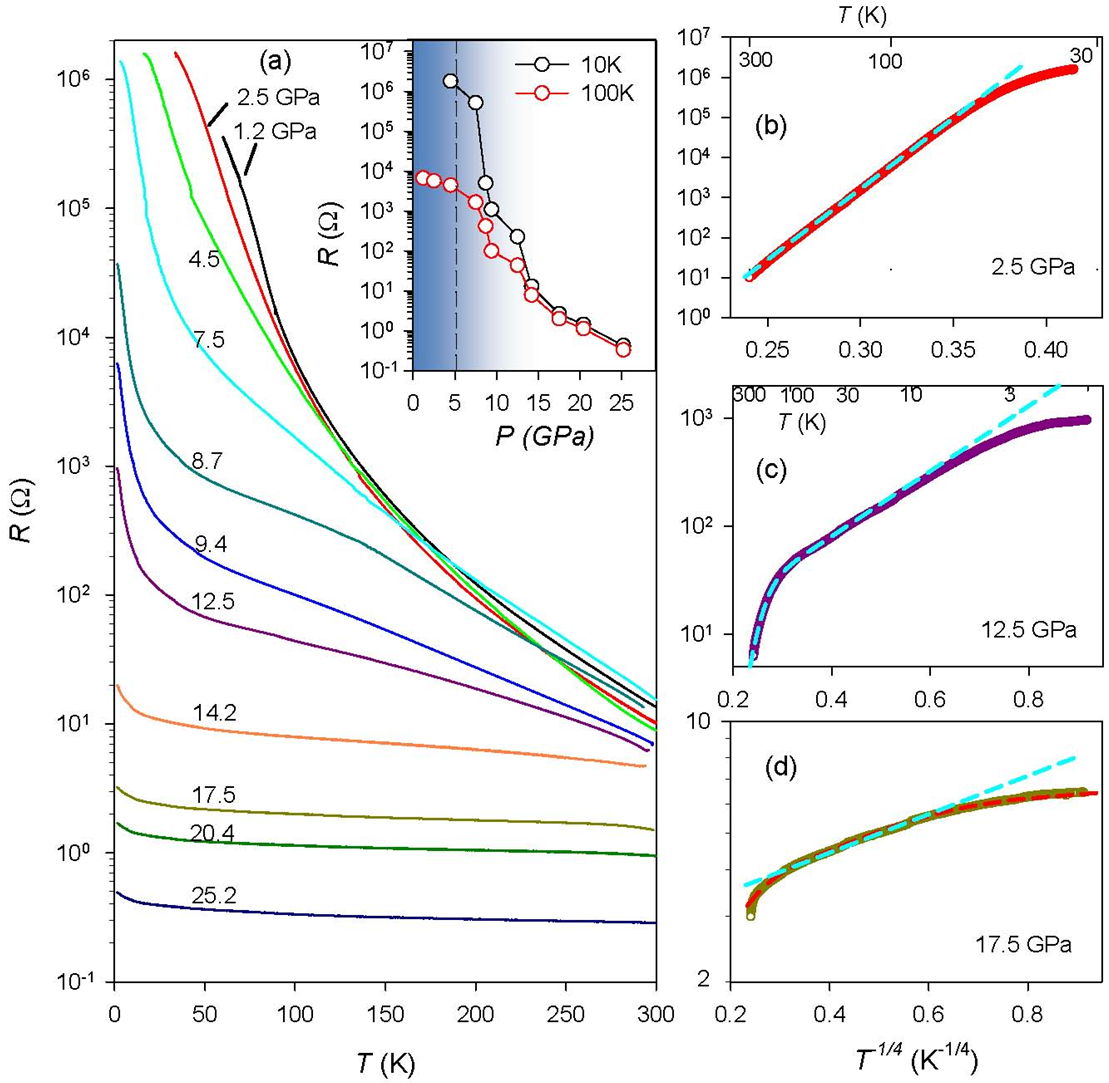}
\caption{\small (a) Semi-log plot of resistance variation with temperature, \textit{R} (\textit{T}), on the pressure cycled Cu$_2$IrO$_3$ sample at various pressures. Inset shows change in sample resistance with pressure at 10 K and 100 K, resistance decreases rapidly above $\sim$5 GPa. Resistance plotted as a function of $T^{-1/4}$, and fitted curves (cyan dashed lines) with (b) 3D Mott VRH conduction term below 4.5 GPa, (c) combination of two VRH terms with two distinct $T_0$ parameters corresponding to low \textit{P} and high $P$ phases in the $P$ range $\sim$5-15 GPa and (d) the VRH conduction of high $P$ phase above 15 GPa. For $P >$ 15 GPa (d), a better fit (red dashed line) is obtained by adding a metallic conduction component with the VRH term (see main text).}
\label{Fig6}
\end{figure}

Figure~\ref{Fig6}(a) shows the resistance variation with temperature (1.4-300 K) on the pressure cycled Cu$_2$IrO$_3$ sample (as discussed above) at various pressures. The $R$($T$) curves remain almost unchanged up to 4.5 GPa, with systematic decrease in $R$ at higher pressures for $T <$ 50 K. In the low pressure range $R$($T$) obeys 3D Mott variable range hopping (VRH) conduction $\frac{R}{R_0} = exp\left[\left(\frac{T_0}{T}\right)^{1/4}\right]$ in the temperature range 50-300 K [see Fig.~\ref{Fig6}(b)]. At low $T$ ($<$ 50 K), a significantly reduced resistance deviating from the VRH fit can be noticed, indicating suppressed scattering channels for the charge carriers, which is yet to be understood. With increasing pressure, the increased Ir-Ir overlap enhances the carrier hopping resulting in further reduction in low temperature resistance. The Mott gap estimated from the high $T$ Arrhenius fit ($\sim$0.3 eV), remains nearly unchanged up to 4.5 GPa.~\cite{suppl}

\begin{figure}
\centering
\includegraphics[width=100mm,clip]{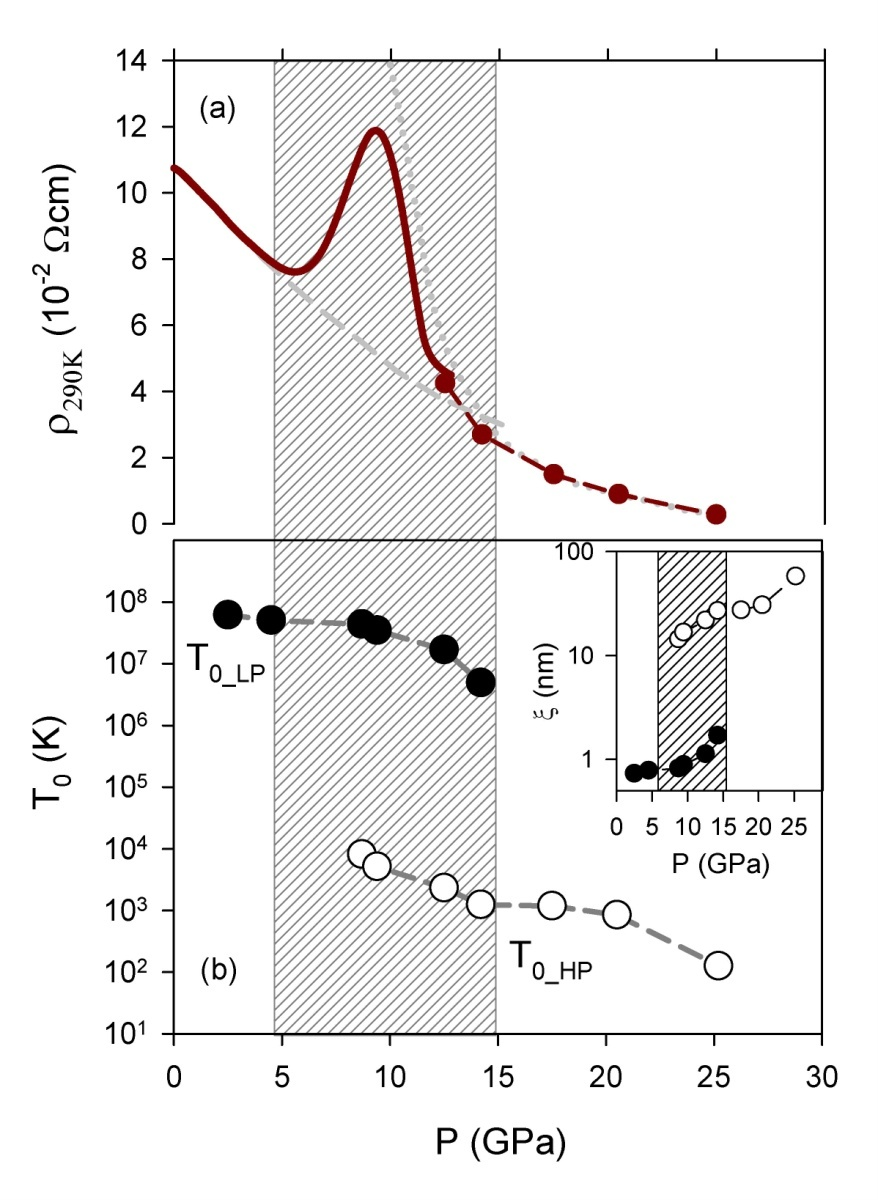}
\caption{\small (a) Room temperature resistivity of pressure cycled polycrystalline Cu$_2$IrO$_3$ sample as a function of pressure. Resistance was measured while increasing $P$ in continuous mode at a slow rate up to 13 GPa and then in larger steps up to 25 GPa. The peak shaped $\rho$($P$) behaviour in the mixed phase region can be decomposed into two $\rho$($P$) curves (shown by dashed and dotted lines) of the two phases. (b) Variation of $T_0$ parameters of the 3D Mott VRH conduction for two phases across the transition. Inset shows that the carrier localization lengths increase with increasing pressure in both phases (assuming unaffected DOS at $E_f$ at high pressures).}
\label{Fig7}
\end{figure}

At higher pressures, a conspicuous change in the $R$($T$) can be noticed. Upon increasing pressure to 17.5 GPa, the resistance at 10 K decreases by six orders of magnitude, indicating dramatic change in its electronic structure [see inset in Fig.~\ref{Fig6}(a)]. Although the resistivity at this pressure reduces to $\sim$10 m$\Omega$. cm, typical for a poor metal, a non-metallic $T$-dependence ($\frac{d\Omega}{dT} <$ 0) is observed over the entire $T$-range. With further increase of pressure, the resistance monotonically decreases with non-metallic $R$($T$) persisting up to 25 GPa, the highest pressure of this investigation. A similar effect of pressure on resistance has earlier been reported on its parent compound Na$_2$IrO$_3$~\cite{Xi2018,Layek2020}, where rapid collapse of resistance at 4 GPa was identified with possible subtle structural modification. In Cu$_2$IrO$_3$, $R$($T$) above 4.5 GPa does not follow either Arrhenius or Mott VRH conduction behaviour. Because of a clear structural transition in Cu$_2$IrO$_3$ with mixed phases in 4 to 15 GPa, we attempted to explain the temperature dependence of $R$ above 4.5 GPa using two phase model.~\cite{suppl} A fit with Arrhenius conduction by invoking defect states~\cite{Schnelle2021} or a combination of Arrhenius and 3D VRH conduction were not successful. However, a combination of two 3D Mott VRH conduction terms with distinct energy scales,
\begin{equation}
\frac{1}{\rho(T)} = \sigma(T) = f\sigma_0 \; exp\left[-{\left(\frac{T_0}{T}\right)}^{\frac{1}{4}}\right] + (1-f){\sigma_0}^{'} \; exp\left[-{\left(\frac{{T_0}^{'}}{T}\right)}^{\frac{1}{4}}\right]
\label{eq_VRH}
\end{equation}
($f$ being the low-pressure phase fraction as determined from our XRD results), is seen to fit the $R$($T$) curves well down to $\sim$10 K [Fig.~\ref{Fig6}(c)]. This can be understood in terms of distinct VRH energy scales $T_0 = T_{0_{LP}}$ and ${T_0}^{'} = T_{0_{HP}}$ of the low $P$ and high $P$ phases and their systematic $P$ dependence [Fig.~\ref{Fig7}(b)].

As the structural transition is completed at 15 GPa and only the high-pressure triclinic phase (with heavily distorted honeycomb lattice) is present above this pressure, $R$($T$) plots above 15 GPa can be fit reasonably well with the second term of Eq.~\ref{eq_VRH} [see the cyan dashed line in Fig.~\ref{Fig6}(d)]. However, in this pressure region, $R$($T$) fit is found to improve significantly [see red dashed line in Fig.\ref{Fig6}(d) and Fig.~\textcolor{blue}{S8}]~\cite{suppl} by including metallic conduction term along with the 3D Mott VRH conduction (with energy scale $T_{0_{HP}}$),
\begin{equation}
\sigma(T) = \sigma_{0_{HP}} exp\left[-\left(\frac{T_{0_{HP}}}{T}\right)^{\frac{1}{4}}\right] + {\sigma_0}^{m} \left[\frac{1}{1 + AT + BT^2}\right]
\label{eq_VRH2}
\end{equation}
While the VRH conduction dominates at high temperature ($T >$ 50 K), the metallic conduction takes over at lower temperatures. Note that with much reduced characteristic energy scale ($T_{0_{HP}}$) of the VRH term, the high $P$ phase is already on the verge of metallization [$E_a \: \sim$10 meV, see Fig.~\textcolor{blue}{S7}(c)]. The inclusion of additional metallic conductivity can perhaps be justified due to the presence of Cu$^{2+}$ defects in the honeycomb layer, adding excess density of states at the Fermi level, which gets noticeable in the high $P$ phase~\cite{Kenney2019}. It is pertinent here to note that semimetals often exhibit mixed carrier (non-metallic and metallic) conduction~\cite{Xu2014,Pavlosiuk2015}.

In Fig.~\ref{Fig7} we summarize the results of our resistance measurements. Figure~\ref{Fig7}(a) shows the plot of room temperature resistivity $\rho$($P$) up to 25 GPa. Upon increasing $P$ up to $\sim$5 GPa (primarily in the low $P$ phase), the $\rho$ decreases monotonically. With the emergence of triclinic phase above this pressure, $\rho$ first increases rapidly up to 9 GPa above which its decreases sharply. The increase of $\rho$ above 5 GPa can be due to increased phase fraction of the high $P$ phase (nano domains) having large resistivity. As the domains grow in size, $\rho$ decreases rapidly above 9 GPa. The interplay between domain size, phase fraction, and intrinsic resistivity of the high $P$ phase may result in this non-monotonic $\rho$($P$) in the mixed phase region. 

In Fig.~\ref{Fig7}(b) are plotted the VRH $T_0$ values (in log scale) as a function of $P$. The characteristic energies of the two phases ($T_{0_{LP}}$ and $T_{0_{HP}}$) decrease with pressure. Inset of Fig.~\ref{Fig7}(b) shows the variation of the Mott localization length ($\xi$) in both phases as a function of pressure using the equation $T_0 = \frac{21.2}{k_B \xi^3 g}$, $g$ being the non-vanishing density of states at the Fermi level. Mott localization length ($\xi$) has been estimated assuming $g$ to be similar to that of ambient pressure Na$_2$IrO$_3$ ($\sim$10$^{-5}$ eV$^{-1}$. \AA$^{-3}$)~\cite{Rodriguez2020}. In the low $P$ monoclinic phase, $\xi$ = 0.7 nm and $\xi/r \; \sim$2, $r$ being the Ir-Ir intersite distance, indicating significant intersite hopping. For this phase, $\xi$ starts increasing above 5 GPa. A much lower $T_{0_{HP}}$ of the high-pressure triclinic phase gives $\xi$ = 12 nm at 9 GPa, assuming an unchanged DOS at the Fermi level (as the disorder states may remain unaffected in the high $P$ phase). An order of magnitude increase of the localization length clearly indicates the high $P$ phase to be of more itinerant character. However, a significant increase of DOS may also be responsible for the reduction of $T_{0_{HP}}$. While a detailed band structure calculation will certainly help to identify its origin, a much reduced $T_{0_{HP}}$ clearly indicates the high $P$ phase to be close to a localized-itinerant crossover.

\subsection{Density Functional Theoretical Calculations}

\subsubsection{\bf Ground state at $P$ = 0 GPa}

$\rm Cu_2 IrO_3$ at ambient conditions occur in the monoclinic $C2/c$ phase~\cite{Abramchuk2017}  with 8 formula-units (f. u.) per conventional unit cell. In a recent study Fabbris \textit{et al.}~\cite{Fabbris2021} has showed another structure ($P2_1/c$ with 4 f.u. per unit cell) to be also consistent with the experimental XRD pattern and lower in energy (7 meV/f.u.) than the $C2/c$ structure reported earlier. We have performed complete structural (atomic positions + lattice parameters) relaxation calculations for both the structures ($C2/c$ and $P2_1/c$), and confirm that $P2_1/c$ is the ground state structure, 13 eV/f.u. lower in energy than $C2/c$ (consistent with Ref.~[\onlinecite{Fabbris2021}]). Our estimated lattice parameters of the ground state structure $P2_1/c$ ($a$ = 6.00 $\si{\angstrom}$, $b$ = 9.49 $\si{\angstrom}$, $c$ = 5.43 $\si{\angstrom}$, $\alpha$ = 89.9, $\beta$ = 107.9, $\gamma$ = 90.0) agree well with those reported earlier~\cite{Fabbris2021} (see SI~\cite{suppl} for details).

\begin{figure}[H]
    \centering
    \includegraphics[width=140mm,clip]{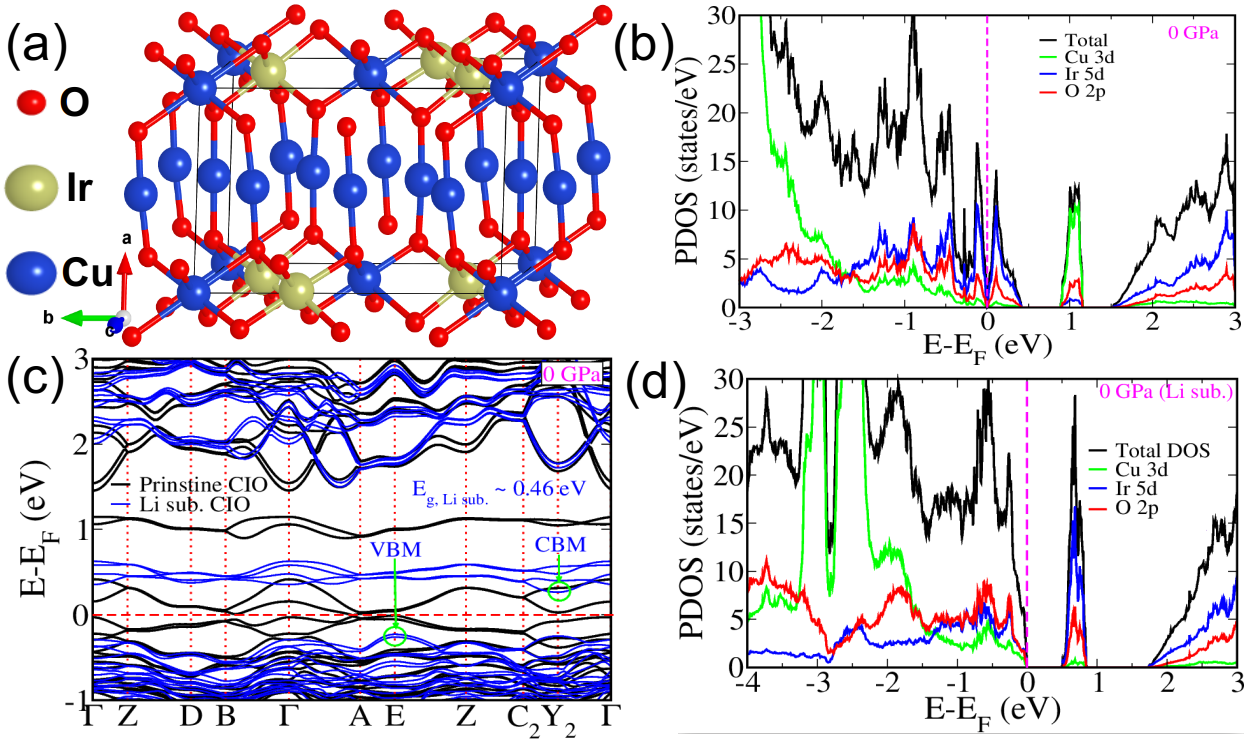}
    \caption{\label{fig:cio} \small (a) Conventional unit cell of $\rm Cu_2IrO_3$ in $P2_1/c$ structure with 2D honeycomb layers are formed by Cu-Ir atoms. (b) Orbital projected electronic density of state of Cu$_2$IrO$_3$ at 0 GPa shows metallic nature with a pseudo gap which is due to intrinsic mix valence disorder. (c) Electronic structure (see Fig.~\textcolor{blue}{S10} for high symmetry path and Brillouin zone) of pristine $\rm Cu_2IrO_3$ (black) and Li-substituted $\rm Cu_2IrO_3$ (blue) at 0 GPa shows Li-substitution at honeycomb layer of $\rm Cu_2IrO_3$ (Cu of 2b site is subsitituted with Li) opens up a band gap of 0.46 eV. (d) Orbital projected electronic density of state of Li-substituted $\rm Cu_2IrO_3$ at 0 GPa shows insulating behavior similar to Kenney \textit{et al}.~\cite{Kenney2019}}
\end{figure}

Experimentally, an electronic energy bandgap  $\sim$ 0.3 eV was observed for $\rm Cu_2 IrO_3$ in $P2_1/c$ phase at $P$ = 0 GPa. In contrast, calculated electronic band structure and orbital projected electronic density of states (PDOS) at $P$ = 0 GPa (see Fig.~\ref{fig:cio}) show that $\rm Cu_2 IrO_3$ is metallic with a pseudo gap. It is argued that metallicity arises from the mixed valence disorder in honeycomb layer formed by Cu-Ir (Cu in $\rm Cu^{+2/+1}$ and Ir in  $\rm Ir^{+4/+3}$ state) atoms~\cite{Choi2019,Kenney2019}. At $P$ = 0 GPa, $\rm Cu_2 IrO_3$ can have an insulating ground state only if Ir (with SOC and onsite electronic correlation) is in +4 oxidation state (see Ref.~[\onlinecite{PhysRevLett.101.076402}] for details), which removes the mixed valence disorder. To simulate Ir in $\rm Ir^{+4}$ we substituted Cu atoms with Li (Ir and Li have comparable atomic radius 132 nm and 128 nm, respectively; see Fig.~\textcolor{blue}{S12} of SI~\cite{suppl}) in the honeycomb layer of the optimized structure of Cu$_2$IrO$_3$~\cite{Kenney2019}. Unlike Cu, Li exists only in $\rm Li^{+1}$ oxidation state which enforces $+4$ oxidation state of Ir avoiding the possibility of mix valence disorder in the honeycomb layer and thus, opens up a band gap of $\sim$0.46 eV comparable to the experimental band gap (0.3 eV).
		
\subsubsection{\bf Dimerization of Ir-Ir bonds at high pressures}

At ambient conditions $\rm Cu_2IrO_3$ exists in $P2_1/c$ phase consisting of 2D honeycomb layers of Ir-Cu atoms, which are connected by Cu-O dumbbells (see Fig.~\ref{fig:cio} and Fig.~\textcolor{blue}{S13} of SI~\cite{suppl}). Our experimental XRD pattern suggests significant bond dimerization of Ir-Ir bond in 2D honeycomb layers in the high-pressure triclinic $P\bar{1}$ phase. In unconstrained (full cell) structural optimization, the high-pressure $P2_1/c$ structure relaxes to $P\bar{1}$ phase which signifies a second order phase transition (transition not involving discontinuous volume change). We could not simulate evolution of difference in enthalpies between $P2_1/c$ and $P\bar{1}$ structures at high pressure, as $P2_1/c$ relaxes to $P\bar{1}$ structure. Our estimates of lattice parameters and Ir-Ir bond lengths at $P$ = 7 GPa reveal dimerization of Ir-Ir bonds. However, calculated high-pressure $P\bar{1}$ phase differ from experimentally observed phase at high pressures (see SI~\cite{suppl}). While resistivity measurements imply that electronic bandgap of $\rm Cu_2IrO_3$ shrinks with pressure, our calculated PDOS of $\rm Cu_2IrO_3$ shows opening of an electronic band gap upon structural transition to $P\bar{1}$ phase (see Fig.~\ref{fig:cio-dos} for pressure evolution of PDOS), similar to earlier work of Fabbaris \textit{et al}.~\cite{Fabbris2021} On the other hand, calculated PDOS for Li substituted Cu$_2$IrO$_3$ shows a band gap of $\sim$ 0.46 $eV$ at 0 $GPa$, which remains almost unchanged with increase in pressure till 12 GPa (see Fig.~\textcolor{blue}{S12} of SI~\cite{suppl}). Difference in pressure dependence of calculated band gap with experimental results could be due to the presence of high degree of dimerization and distortions in the high-pressure structure of $\rm Cu_2IrO_3$, which is absent in our optimized high-pressure phase. 
 
\subsubsection{\bf Pressure dependent Raman spectra}

$\Gamma$-point phonon frequencies (calculated without inclusion of SOC) at $P$ = 0 GPa in the $P2_1/c$ phase (see Table~\textcolor{blue}{S8} of SI~\cite{suppl} for details) confirm its dynamical stability of the low $P$ phase. We calculated $P$-dependence of Raman active phonon mode frequencies with pressure (see Fig.~\ref{fig:cio-raman}) which qualitatively agree with experiment. As our calculation could not capture experimentally observed high-pressure $P\bar{1}$ phase, our estimates of $\frac{d\omega}{dP}$ does not match with experimentally observed values.

\section{Discussions}

In the stability range of the monoclinic structure ($C2/c$) of Cu$_2$IrO$_3$ (for $P <$ 4 GPa), application of hydrostatic pressure does not reduce the inter-honeycomb separation, whereas intra-layer Ir-Ir bond length decreases by more than 1\%. This reduction has no effect on the Ir-O-Ir angles in Ir hexagons, but brings the IrO$_6$ octahedra towards ideal $O_h$ symmetry by elongating along the trigonal symmetry axis ($\perp$ \textit{ab} plane)~\cite{suppl}. This can possibly increase Kitaev exchange ($K$) with respect to Heisenberg exchange~\cite{Kim2014,Winter2016}. Although an increased $d$-$d$ overlap and slightly deviated Ir-Ir-Ir angles from the ideal 120$^{\circ}$ may increase Heisenberg exchange $J$ (nearest neighbour and beyond), the enhanced effective $K/J$ exchange ratio is expected to drive the system towards the Kitaev limit at higher pressures~\cite{Yadav2018_PRB}. This is supported by our observed systematic increase with pressure of the CW temperature in the \textit{dc} magnetic susceptibility as well as the dynamical Raman susceptibility [Fig.~\ref{Fig4}(c)]. The decrease of room temperature resistivity with increasing pressure in the low $P$ phase is due to increased Ir-Ir hopping as a result of the above structural modifications. In this pressure range, $R$($T$) follows a 3D Mott VRH carrier conduction with characteristic energy scale $T_{0_{LP}} \; \sim$ 5$\times$10$^7$ K, in agreement with other iridate compounds~\cite{Layek2020,Rodriguez2020}, that also decreases only marginally with increasing pressure [Fig.~\ref{Fig7}(b)].
  
At higher pressures ($P >$ 4 GPa), the system undergoes structural transition into a triclinic structure ($P\bar{1}$) with heavily distorted honeycomb lattice (with one short Ir-Ir bond length, $\sim$2.41 \AA) and with $\sim$8\% decrease in inter-layer separation. As the above Ir-Ir bond length is shorter than that in iridium metal, a highly covalent bond formation is expected resulting in Ir-Ir dimerization. Our XRD analysis indicate that we do not observe intermediate $\alpha-P\bar{1}$ triclinic phase as reported by Ref.~[\onlinecite{Fabbris2021,Jin2022}] above 7 GPa. Rather, our observed XRD pattern of high-pressure phase agrees with the previously reported high-pressure pattern after the second transition above $\sim$11-15 GPa ($\alpha'-P\bar{1}$). Results of our resistance measurements show that the carrier conduction in the high $P$ triclinic phase is of Mott VRH type with much reduced energy scale ($T_{0_{HP}}$) (weak localization) compared to the low $P$ phase, which progressively decreases on increasing pressure, thus approaching itinerant character (due to either increased localization length or enhanced DOS at the Fermi level). This can be understood in terms of band broadening at higher pressures as well as Ir-Ir covalent bond formation~\cite{Xi2018,Wang2018}. In Li$_2$IrO$_3$ and Na$_2$IrO$_3$, a similar reduced resistivity has been observed at a much higher pressure (above 45 GPa). As the high-pressure phase fraction is low ($<$ 50\%) up to $\sim$9 GPa, our attempt to measure magnetic susceptibility of the high-pressure phase has not been successful. Investigation of possible ordering of bond singlets in the pure high-pressure phase suggests a need to do magnetic measurements above 15 GPa.

At further higher pressures ($P >$ 15 GPa), $R$($T$) shows persistent non-metallic behaviour as seen in similar layered honeycomb compounds, attributed to the crossover of Mott insulator to itinerant behaviour due to formation of additional states at the Fermi level~\cite{Xi2018,Wang2018}. Recently, Jin \textit{et al}.~\cite{Jin2022} have reported insulator to metal transition above 13.8 GPa where a metallic behaviour is observed with a resistance minimum at $\sim$15 K. Although the results of the present study on polycrystalline Cu$_2$IrO$_3$ show persistent non-metallic behaviour up to 25 GPa, detailed $R$($T$) analysis indicates possible metallization in the higher pressure triclinic structure. The different pressure dependence of $R$($T$) curves as compared to that reported by Ref.~[\onlinecite{Jin2022}] may be attributed to different levels of Cu$^+$/Cu$^{2+}$ disorder in the Cu$_2$IrO$_3$ samples. The presence of mixed carrier conduction in the high-pressure triclinic phase can also be due to large pressure inhomogeneity over the sample area that demands further investigations. While persistent non-metallic state is common in layered perovskite iridates~\cite{Chen2020}, Sr$_3$Ir$_2$O$_7$ exhibits pressure-induced anisotropic metallization~\cite{Ding2016}. A non-metallic $T$-dependence with significantly low resistivity can originate from such anisotropic metallization in the high-pressure phase of polycrystalline Cu$_2$IrO$_3$ (with possible metallization only in honeycomb layer). Exploration in this regard requires single crystal Cu$_2$IrO$_3$ (not yet available), but will be of immense help for better understanding of layered honeycomb iridates in general.

In summary, quasi-2D layered honeycomb Kitaev candidate Cu$_2$IrO$_3$ undergoes pressure-induced structural phase transition (monoclinic $C2/c$ $\rightarrow$ triclinic $P\bar{1}$) with a broad mixed phase pressure range ($\sim$4 to 15 GPa). The triclinic phase consists of heavily distorted honeycomb lattice with Ir-Ir dimer formation and a collapsed interlayer separation. In the stability range of the low-pressure monoclinic phase, structural evolution with hydrostatic pressure brings the system closer to Kitaev limit, as indicated by the enhanced CW temperature from magnetic susceptibility, non-emergence of long-range magnetic order and enhanced dynamic Raman susceptibility. High-pressure resistance measurements up to 25 GPa show resilient non-metallic $R$($T$) behaviour with drastically reduced resistivity in the high-pressure phase. A Mott variable-range-hopping conduction with much reduced $T_0$ indicates that the high-pressure phase is close to the localized-itinerant border. Using first-principles calculations, we presented a pressure induced structural transition in $\rm Cu_2IrO_3$ from $P2_1/c$ phase to $P\bar{1}$ phase around 7 GPa. We find Ir-Ir bonds in 2d honeycomb layer get dimerized in $P\bar{1}$ phase which is in line with our experiments. Calculated high-pressure $P\bar{1}$ phase of $\rm Cu_2IrO_3$ differs from experimentally observed high-pressure phase, (see Table~\textcolor{blue}{S3} - \textcolor{blue}{S7} in SI~\cite{suppl} for details). From pressure dependent resistivity measurements, we observe decrease in electronic band gap with increase of pressure. In contrast evolution of calculated electronic energy band gap exhibits  an opposite trend, which is due to underestimation of highly dimerized Ir-Ir bond length. Our calculated $P$-dependence of Raman-active phonon modes qualitatively agree with the experimental Raman results.

\section{Acknowledgements}

{AKS thanks Nanomission Council and the Year of Science professorship of DST for financial support. High-pressure PXRD investigations have been performed at Elettra synchrotron, Trieste under the proposal ID 20195289. AS is thankful to JNCASR and DST (GoI) for providing fellowship and computing facilities. UVW acknowledges support from J. C. Bose National Fellowship of SERB-DST, Govt. of India.}\\

\noindent{\color{blue}{$^{\dagger}$}} Both authors contributed equally to this work.\\
{\color{blue}{$^{\#}$}} Part of this work was conducted by PM during her visit to Bhabha Atomic Research Centre, Trombay, Mumbai 400085, India.

\newpage

\section*{Figures and Tables}

\begin{figure}[H]
    \centering
    \includegraphics[width=140mm,clip]{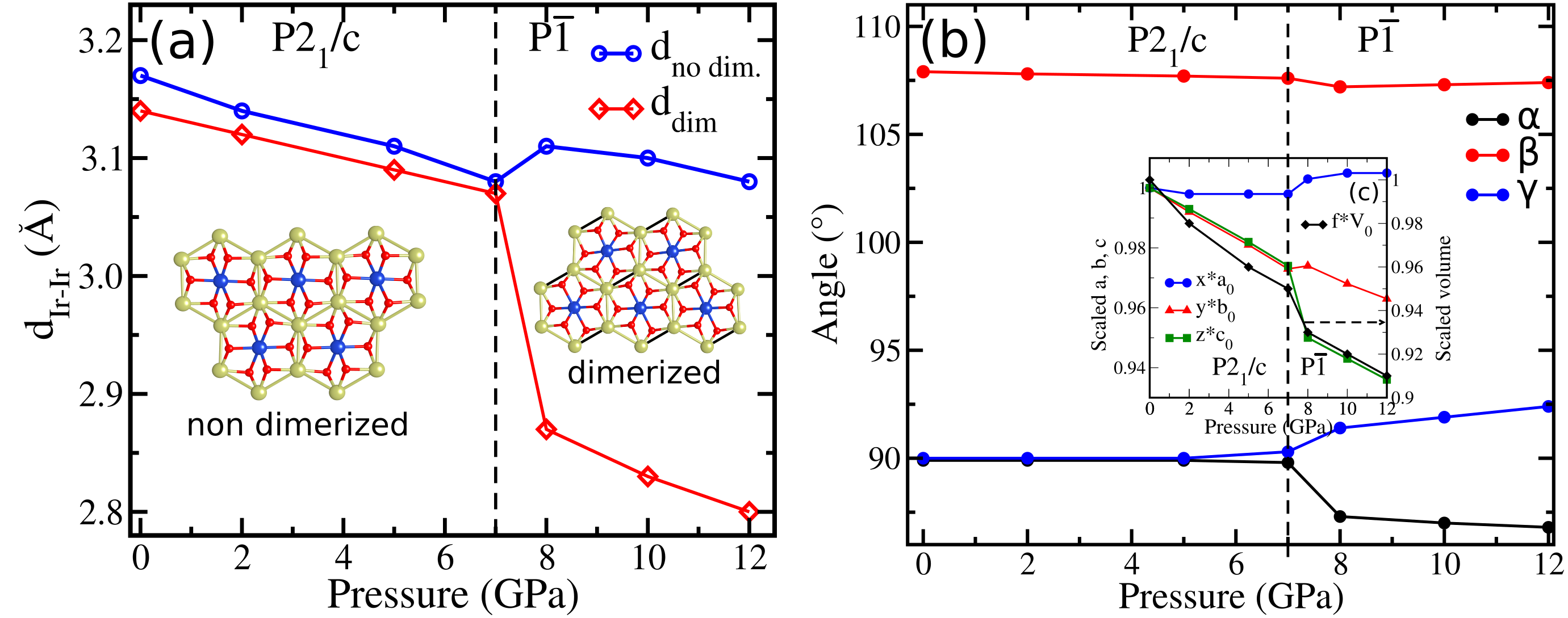}
    \caption{\label{fig:cio-dim} \small (a) Evolution of Ir-Ir bond lengths in the honeycomb layer of $\rm Cu_2IrO_3$ shows dimerization of some Ir-Ir bond with pressure after the transition to $P\bar{1}$ at $P \sim$7 GPa. (b) Evolution of optimized lattice parameters $\alpha$, $\beta$, $\gamma$ of Cu$_2$IrO$_3$ with pressure confirms a phase transition from $P2_1/c$ to $P\bar{1}$ around 7 GPa. Inset shows evolution of scaled lattice parameters (wrt ambient $a_0,\;b_0,\;c_0,\;V_0$) $a$, $b$, $c$ and volume $V$ with pressure showing structural transition at 7 GPa. Small change in volume during transition and relaxing of $P2_1/c$ phase at higher pressures established this transition as second order phase transition.}
\end{figure}

\begin{figure}[H]
    \centering
    \includegraphics[width=160mm,clip]{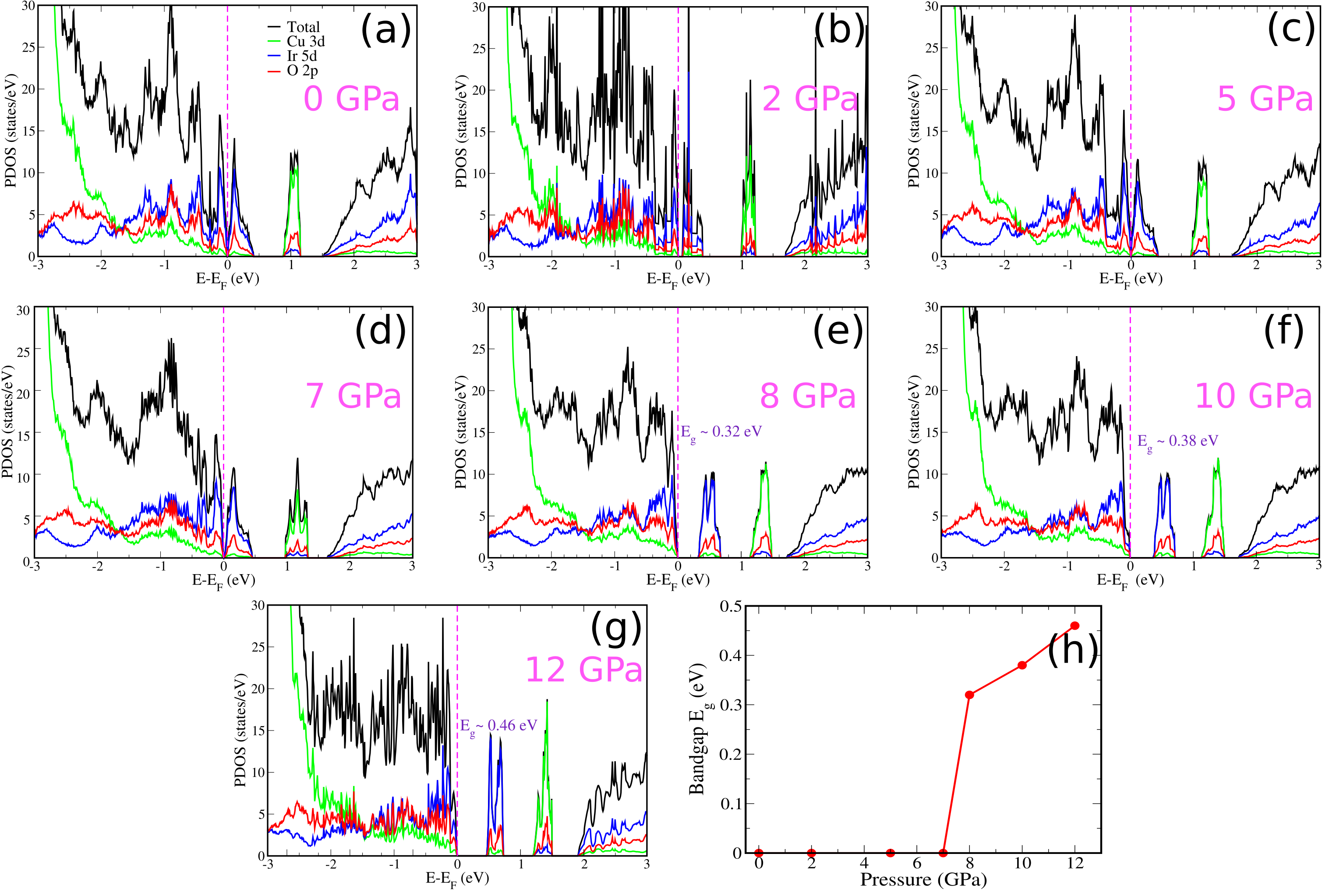}
    \caption{\label{fig:cio-dos} \small (a) Evolution of orbital projected electronic density of states of pristine $\rm Cu_2IrO_3$ with pressures $P$ = (a) 0 GPa, (b) 2 GPa, (c) 5 GPa, (d) 7 GPa, (e) 8 GPa, (f) 10 GPa (g) 12 GPa, and (h) pressure evolution of electronic band gap. (a) - (f) show that before dimerization of Ir-Ir bond, Cu$_2$IrO$_3$ is metallic with a pseudo-gap at Fermi level and states near the Fermi level dominated by Ir $5d$ orbitals. After the transition to $P\bar{1}$ phase (around 7 GPa), bandgap opens up and increases with pressure which is consistent with earlier work~\cite{Fabbris2021}.}
\end{figure}

\begin{figure}[H]
    \centering
    \includegraphics[width=100mm,clip]{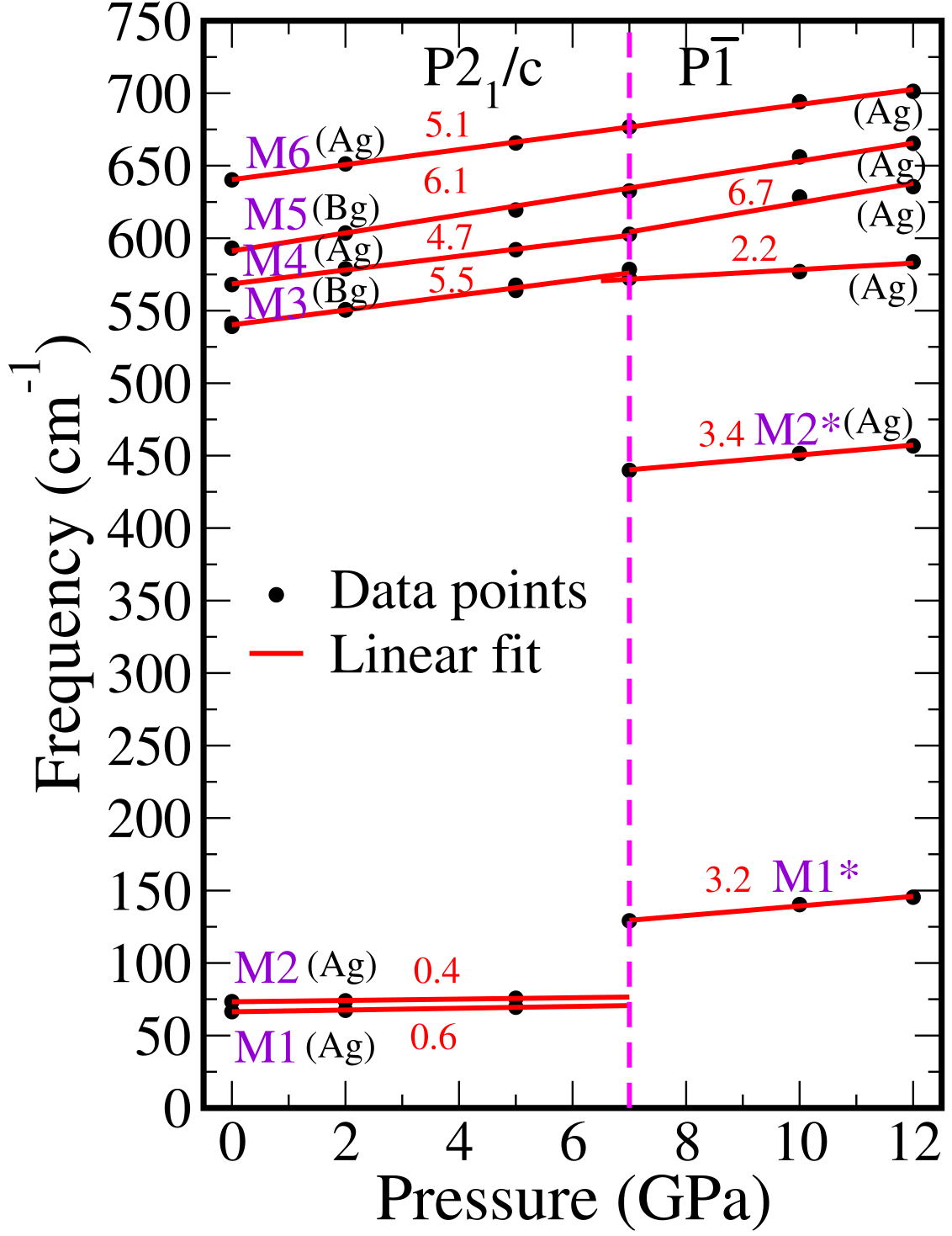}
    \caption{\label{fig:cio-raman} \small Pressure variation of some of the calculated $\Gamma$-point Raman-active phonon mode frequencies in the low-pressure $P2_1/c$ phase and high-pressure $P\bar{1}$ phase. Numbers near the lines denote slopes $\frac{d\omega}{dP}$ (mode symmetries are indicated in parentheses).}
\end{figure}




\end{document}